# Decision-making processes underlying pedestrian behaviours at signalised crossings: Part 2. Do pedestrians show cultural herding behaviour ?

Marie Pelé[1], Jean-Louis Deneubourg[2], Cédric Sueur[3]

[1] Ethobiosciences, Research and Consultancy Agency in Animal Well-Being and Behaviour, Strasbourg, France

[2] Unit of Social Ecology, Université libre de Bruxelles, Brussels, Belgium

[3] Université de Strasbourg, CNRS, IPHC UMR 7178, F-67000 Strasbourg, France

Corresponding author: marie.pele@iphc.cnrs.fr, 0033(0)88107453, IPHC UMR 7178, 23 rue Becquerel F-67000 Strasbourg, France

**Abstract:** Followership is generally defined as a strategy that evolved to solve social coordination problems, and particularly those involved in group movement. Followership behaviour is particularly interesting in the context of road-crossing behaviour because it involves other principles such as risk-taking and evaluating the value of social information. This study sought to identify the cognitive mechanisms underlying decision-making by pedestrians who follow another person across the road at the green or at the red light in two different countries (France and Japan). We used agent-based modelling to simulate the road-crossing behaviours of pedestrians. This study showed that modelling is a reliable means to test different hypotheses and find the exact processes underlying decision-making when crossing the road. We found that two processes suffice to simulate pedestrian behaviours. Importantly, the study revealed differences between the two nationalities and between sexes in the decision to follow and cross at the green and at the red light. Japanese pedestrians are particularly attentive to the number of already departed pedestrians and the number of waiting pedestrians at the red light, whilst their French counterparts only consider the number of pedestrians that have already stepped off the kerb, thus showing the strong conformism of Japanese people. Finally, the simulations are revealed to be similar to observations, not only for the departure latencies but also for the number of crossing pedestrians and the rates of illegal crossings. The conclusion suggests new solutions for safety in transportation research.

**Introduction**

"We are discreet sheep; we wait to see how the drove is going, and then go with the drove" [1]

The fundamentally gregarious nature of human beings leads them to follow simple rules in the context of collective phenomena such as panic, pedestrian traffic flow and crowd coordination [2–5]. Indeed, who can resist following a person who is running in a specific direction, even without knowing if it is justifiable? During such collective events, it has often been shown that a number of individuals take the initiative, then others follow [6,7]. This followership is generally defined as a strategy "that evolved for solving social coordination problems in ancestral environments, including in particular the problems of group movement, intragroup peacekeeping and intergroup competition" [8,9]. This followership behaviour is particularly interesting in the context of road-crossing behaviour because it involves other principles such as risk taking [10,11] and evaluating the value of social information [12,13].

Walking in the street is a necessary daily behaviour that is generally considered to be safe in the light of the number of times people cross the road without getting struck by a vehicle. Yet crossing the road causes the largest number of pedestrian accidents and the most severe injuries [14]. Indeed, circumstances such as the urban and social environment can make crossing the road a high-risk behaviour. One of these circumstances is the misuse of social information [12,15]. Indeed, social information, i.e. how to trust information gained from people surrounding us, is an important topic in the context of human collective phenomena and followership [16–19]. In the case of road crossing behaviours at signalled crossings, trusting wrong or unreliable information and following someone crossing at the red light without checking the light colour by ourselves (i.e., personal information) leads to an increased risk of accidents and injuries [20]. Indeed, when crossing the road at a signalised crossing, it is rare for pedestrians to look at the light colour when other pedestrians have already started crossing: they just follow. This probability of following depends on many factors such as gender or spatial proximity [12] and is governed by an amplification process also called information cascade or mimetic process [12,15,21,22]. An information cascade [21,23] occurs at signalised crossings when pedestrians observe the actions of others and then do likewise, despite possible contradictions with their personal information. The mimetic process is quite similar, except that in this case the probability of crossing increases with the number of pedestrians that have already started crossing according to what we call a mimetic coefficient [22]: the higher the number of pedestrians that have already started crossing, the more likely it is that the remaining individuals will follow them, even if the former took the wrong decision (i.e. they crossed at the red light). However, some studies also suggest that behaviours and reactions vary according to the socio-demographic traits of individuals.

For example, men show more high-risk behaviours than women, who display more compliance and conformity [24–26]. In the same way, the rate of rule-breaking differs according to the country in which the pedestrian study is done, and this is not due to a difference in the risk of car-pedestrian accidents but to social conventions [27] and cultural differences between societies that are conformist or individualistic [28–30].

In this study, we wanted to highlight the cognitive mechanisms underlying decision-making in pedestrians who follow another person who is crossing at the green or red light. Although many studies have attempted to understand which factors influence the number of rule-breaking at road-crossings, very few have focused on the real decision-making processes of pedestrians facing the situation of following other pedestrians at a signalised crossing and how the decision to cross is affected by changes in the number of pedestrians that have started crossing [31,32]. We compared the behaviour according to the country (France and Japan), but also differentiated between crossings according to the gender of individuals. Mathematical (equations) and survival analyses were used to better understand the cognitive mechanisms underlying the decision-making process [22,33]. However, as these types of analyses may only provide a range of values to explain behaviours rather than a precise figure of the considered parameter, we also used agent-based models to simulate decision-making processes and followership behaviours [34–36]. The use of agent-based models made it possible to recreate all the variables observed at pedestrian crossings, test them thousands of times and then compare the results with those observed with the simulated data. This kind of analysis is not possible using traditional statistical approaches [36,37].

The study assumed that pedestrians crossing the road after one or several pedestrians had done likewise show an amplification process or a mimetic behaviour, meaning that the higher the number of pedestrians crossing the road, the higher the probability is that others will follow them. We tested this mimetism hypothesis against the null hypothesis of independence, meaning that the probability of crossing the road is not governed by an amplification process but solely by each pedestrian's intrinsic probability of crossing. This suggests that at least two processes underlying pedestrian decisions: their own motivation, or *intrinsic probability*, and the likelihood that they will be influenced by others, i.e. the mimetic process [22,38]. These processes might however vary according to individuals and their socio-demographic traits, which leads us to make several assumptions. First, the intrinsic probability of crossing and the mimetic process should both be lower at the red light compared to the green light because a minority, if not a majority of pedestrians respect the rules and consider risks [39–41]. As far as the country or cultural effect are concerned, French pedestrians are expected to show a higher intrinsic probability to cross at the red light but no difference at the green light, as French people are known to be more individualistic and less conformism than Japanese people. The mimetic process should be different between French and Japanese pedestrians, and mimetism in French citizens is expected to be stronger mimetic process at the red light but lower at the green light compared to their Japanese counterparts. However, as the Japanese are more conformist and are aware of group pressure [28–30], they are expected to consider not only the number of pedestrians that are already crossing, but also the number of pedestrians remaining on the sidewalk, whilst French pedestrians are expected to only take the number of crossing pedestrians into account. This process, which takes both the number of moving and of waiting individuals into account, is a "following the majority" process [23,42] that has also been observed in sheep during decisions to move together [43,44]. Men are shown to take more risks [11,25] and be less concerned by group pressure, and should therefore show higher intrinsic probability than women to cross the road at the red light for instance. Similarly, men are expected to follow the pedestrians that have already started crossing, and are not expected to consider the number of waiting pedestrians, whilst women are more likely to check the behaviours of resting pedestrians as they are more sensitive to group pressure [24,26].

**Material & Methods**

    **a.** Study sites

We observed pedestrian behaviours at three sites in Strasbourg, France and at four sites in Nagoya, Japan. Details of each site are given in Table 1. Pictures of each site are available in [15]. These sites all permitted the observation of collective road crossings involving at least 10 pedestrians at a time. The speed of vehicles on each site was limited to 50kmh$^{-1}$. There was no difference in pedestrian crossing speed between the sites (permutation test for independent samples: maxT=2.22, p=0.168). At some sites, vehicles were allowed to turn left or right despite the green light for pedestrians, but the drivers were aware that crossing pedestrians had priority.

Moreover, turning vehicles travel much slower than vehicles that are heading straight on. However, the driver of an approaching vehicle may be less careful if pedestrians cross at the red light, as he/she has the right of way. The risk to pedestrians is therefore much higher when crossing at the red light. There was no button for pedestrians to trigger the green light at any of the sites studied.

    b.  Data scoring

Data were scored over a six-day period for each site, for 1 h per day during working days, hours and weeks to ensure that data excluded movements generated by tourism, festivals, etc. This scoring duration is sufficiently ample to provide a large dataset [12,32,45]. Video cameras were set up in order to score the light colour and were placed in locations ensuring that crossing pedestrians were visible at all times. Behavioural sampling was used to score the crossing of pedestrians in one direction only, i.e. that recorded by the camera. Pedestrians were not informed about the purpose of the study. As both cities are touristic, pedestrians are accustomed to seeing tourists taking pictures or videos. We did not observe any difference in the way pedestrians behaved when they saw the camera. We did not take any other equipment such as counters or pocket PC in order to avoid influencing pedestrian behaviour. When observation of road-crossing behaviour was hampered by a visual obstacle (i.e. a car or a truck in front of the video camera), we removed this behaviour and the behaviours occurring immediately before and after it from the analyses. We also removed data in which cyclists were among the pedestrians or where tourists were present. Tourists were easily differentiated from citizens, as they were in large groups accompanied by a guide, were dressed differently from citizens and carried specific equipment (guidebook, map, camera, etc.).

c. Research ethics

Our methodological approach solely involved anonymous observations and anonymous data scores. Our protocol followed the ethical guidelines of our institutions (IPHC, Strasbourg, France and PRI, Kyoto University, Japan) and we received ethical approval from these institutions to carry out our study. All data were anonymous, and individuals were given sequential numerical identities according to the time of the road crossing and the arrival/departure order of crossing. Pedestrians had the possibility to be informed about the study by an information medium in their language (Japanese or French). They were also provided with an email address and phone number to contact our institution at a later date if desired. Persons who refused to participate in the study were removed from the data (i.e. we deleted the crossing concerned).

d. Data analysis

This study, focuses solely on following pedestrians (at the green and the red light) and not the first pedestrians to go, which have been described in a first paper, "Part 1" (Pelé et al., submitted). This approach was chosen because the two types of decisions (departing first and following) are underlain by very different processes [9,46,47]. The complete six-hour data set was analysed for each site. We scored the behaviours of following pedestrians when at least two pedestrians crossed the road simultaneously (i.e. when the time between the two departures was lower than the mean road-crossing time indicated in Table 1 for each site).

All indicated times are in hundredths of a second. We decided to limit our analysis to the crossing behaviours of the first 10 pedestrians at the green or at the red light, mainly due to the difficulty of analysing the time and order of crossing when more than 10 pedestrians were crossing the road collectively. Similarly, we only analysed data for pedestrians who were present at the crossing when the light colour changes, or who either decreased their walking speed or stopped as they approached the crossing, as we wanted to assess how specific factors such as the light colour, waiting time, and number of pedestrians influenced their decisions to cross the road.

We scored road crossings for 2568 followers, of which 1839 crossed at the green light and 729 at the red one. Nine hundred and two crossings of followers were scored in France and 1667 in Japan.

For each following pedestrian, we scored the following variables (see [15] for a visual explanation of the different scored variables):

— The light colour when crossing (red or green).

— The departure time, $\Delta T_j$, i.e. the period between the previous light colour change and the moment the pedestrian $j$ starts crossing the road. This variable is positive for pedestrians crossing after the light (for pedestrians) has turned red but negative for pedestrians crossing before the light turns red.

— The departure latency $\Delta T_{j,j-1}$, i.e. the time elapsed between the departure of pedestrian j and previous pedestrian j−1.

— The departure order of pedestrians, where the first pedestrian to leave the kerb is ranked as 1, the second as 2, and so on. Here, we then focused on pedestrians of ranks 2 to n, n being the number of following pedestrians in a crossing event, with a maximum threshold set at n=10.

— The gender (male or female).

— The age, estimated at 10-year intervals from 0–9, 10–19 [ . . . ] to 70–89. However, given the number of data and the analyses we carried out, it was not possible to analyse the effect of age (per interval) on the decision-making processes.

— The country (France or Japan).

— The waiting time, i.e. the time between the moment a pedestrian stops at the light and the moment he/she starts crossing the road.

e. Mathematical analyses

Survival analysis [33,48] was used to study the distributions of departure latencies for all followers. Survival analysis indicates how the ratio of observations decreases from 1 (all observations/data) to 0 (none) according to a response variable. First, curve estimation tests identified which type of function was followed by these distributions, namely linear (meaning that the probability of crossing is time dependent), exponential (the probability of crossing is constant over time) or sigmoid (the probability of crossing depends on a time threshold that is directly correlated to the response variable) (see [11,22,38].

When trying to identify the process underlying following behaviours in pedestrians, we had two hypotheses, namely the independence hypothesis and the mimetism hypothesis [21,22,49]. Explanations of these two different hypotheses are given just below, but they can only be considered once we have evaluated whether the intrinsic possibility of crossing for each pedestrian is constant per time unit. This intrinsic motivation is given by studying the distribution of departure latencies for the first pedestrian who crosses. As this distribution corresponded to an exponential distribution (see Supplementary Information and Table 2), the departure probability of the first pedestrian to cross $\psi_{01}$ is the log gradient of this exponential distribution, that is, the inverse function of the mean departure latency $\Delta_{T1}$ for the first departing individual [38,49,50]:

(1a) $\psi_{01} = \sum_{i=1}^{n} \lambda_i$

where $\Psi_{01}$ is the probability of seeing a first pedestrian crossing the road. We based this probability on the departure latencies for a pedestrian crossing at the green light, but for pedestrians crossing at the red light, we based this probability on waiting time.

$n$ is the numbers of waiting individuals, here n=N being the maximum number of pedestrians we analysed for each crossing, meaning 10.

$\lambda_i$ is the probability of individual $i$ being the first person to cross. Here, we analysed this probability according to the gender and the country (Table 2).

The probability $\lambda_{i,1}$ of the individual i to be the first to cross is therefore

(1b) $\lambda_{i,1} = \frac{\psi_{01}}{n} = \frac{\psi_{01}}{10}$

And (1c) $\psi_{01} = \frac{1}{\Delta_{T1}}$

In a mimetic process where the departure probability is proportional to the number of pedestrians already moving $j$, the probability $\psi_j$ per unit time that one of the $n$ waiting agents became the $j$th following pedestrian is:

(2a) $\psi_j = \left(\lambda + C(j-1)\right)n$

where $C$ is the mimetic coefficient per individual. Here, we analysed this probability depending on the gender and the country (results are given in Table 2).

The departure latency $\Delta T_{j,j-1}$ of the follower $j$ was:

(2b) $\Delta T_{j,j-1} = \frac{1}{(\lambda + C(j-1))(n-(j-1))}$

or

(2c) $\frac{1}{\Delta T_{j,j-1}} = \psi_{j,j-1} = (\lambda + C(j-1))(n-(j-1)) = (\lambda - C)(n+1) + j(2C + Cn - \lambda) - Cj^2$

This is a parabolic law $z+wj+yj^2$ where

$z = (\lambda - C)(n+1)$

$w = (2C + Cn - \lambda)$

$y=C$

Fitting the distribution of departure probability per pedestrian departure rank with a parabolic curve allows us to calculate a range of values for the mimetic coefficient $C$ (See "Supplementary Information" for calculation details and "Results" for the statistical values of parabolic curve estimation tests).

f. Modelling

A modelling approach using multi-agent based models was used to determine whether pedestrians cross according to a mimetic process (mimetism hypothesis) or independently of external influence (independence hypothesis), and to ensure the precise determining of the mimetic coefficient for each gender and country according to the range of values provided by the survival analyses.

We implemented the distribution of the number of waiting pedestrians N in the model. There is no difference in this distribution between the Japanese and French sites (Mann-Whitney test= W=764990, p=0.464, $M_{France}$=12.87±6.14, $M_{Japan}$=12.38±4.59).

At the start of a simulation, all agents ($N$) were at the kerb on the same side of the road (we consider only one side) and had to move to the opposite sidewalk. We implemented the departure probability $\lambda_i$ of each agent, according to his or her gender and his or her country (see Table 2). Pedestrians of the same gender and from the same country all have the same probabilities. The departure probability of the first departing individual, the only individual whose decision to move would not be influenced by the other group members, was identical for the two versions of the model tested here, i.e. the independence hypothesis and the mimetic process hypothesis. We assumed in the model that all agents are aware of the state (waiting or crossing) of all other agents at any given time.

i. Independence hypothesis

The first hypothesis assumed that individuals were independent: the departure probability of a pedestrian was not influenced by the departure of other pedestrians. As a consequence, the individual departure probability remained constant, whatever the rank of the crossing pedestrian. Under this hypothesis, the probability that one of the $n$ waiting agents (e.g. individual $i$) would became a crossing pedestrian $j$ per unit time was $\lambda_i$. According to equations (1a) and (1b), the departure latency of the crossing pedestrian $j$ was the inverse function of the sum of the $\lambda_i$ of n waiting agents:

(3) $\Delta T_{j,j-1} = \frac{1}{\sum_{i=1}^{N} \lambda_i}$

In our case, the probabilities were identical, with $\lambda_1 = \dots = \lambda_N = \lambda$ for pedestrians of same country and same gender.

## ii.   Mimetic process hypothesis

The second hypothesis specified that pedestrians would be influenced by seeing others crossing, thanks to a mimetic process. To test this hypothesis, we added a mimetic coefficient $C$ to the above version of the model (independence hypothesis). The probability per unit time that one of the $n$ waiting agents would become the crossing pedestrian $j$ under the anonymous mimetic hypothesis was obtained from equation (2a) and its departure latency obtained from equation (2b). The range of different mimetic coefficients found in Table 2 was tested. In this version of the model and according to equation (2a), waiting pedestrians had the same probability of crossing but would now be differentiated according to their gender and their country. In equation (2a), the probability that an individual will cross is only influenced by the number of already crossing individuals $j$. This is the most parsimonious hypothesis, which also has the lowest number of parameters [21,44,51]. However, when the number of individuals already moving is not enough to explain the probability of departures, it is necessary to take into account both the number of moving individuals (here, the pedestrians that have already started crossing) and the number of individuals that are not moving (here, the waiting pedestrians) [44]. This has been applied in data analysis for collective movements of sheep [43].

In this last case, the probability to see a pedestrian crossing the road is therefore:

(3)      $\psi_j = \left( \lambda + C \left( \frac{j-1}{n} \right) \right) n = \left( \lambda n + C(j-1) \right)$

If the first rule (equation 2a) did not fit with our observed data, we tested the second rule (equation 3).

The different versions of the model, corresponding to each hypothesis, were implemented in Netlogo 6.0.6 [36,52]. The models we developed are modified versions of an existing model from [38] and is adapted to Netlogo 6.0.6.

The model is stochastic: a number between 0 and 1 was randomly attributed to each waiting pedestrian at each time step ($1\ s^{-100}$). If this number was lower than the theoretical departure probability of each agent, the pedestrian would start to cross; if this number was higher than the theoretical departure probability, the agent did not move and continued to wait. The country, the gender, the pedestrian departure rank and the departure latency of each crossing pedestrian agent were scored for each simulated crossing event. To be consistent with the experimental situation, we stopped a simulation after a specific time threshold when no agent followed the departure of the first or last crossing pedestrian. This threshold depends not only on the light colour but also on the country. When crossing at the green light, the time threshold is defined as the time the light remains green. $8000\ sec^{-100}$ is the maximum green light time between the sites. Knowing that the maximum departure time of a following pedestrian was $976\ sec^{-100}$, this threshold is more than sufficient to avoid committing any errors. At the red light, French and Japanese pedestrians showed two different behaviours (see Part 1, Pelé et al. submitted for details). Indeed, when crossing illegally, French pedestrians crossed throughout the duration of the red light (about $8000\ sec^{-100}$) whilst the Japanese only crossed when the pedestrian light was close to changing from red to green (about $400\ sec^{-100}$). We then set these thresholds for French ($8000\ sec^{-100}$) and Japanese ($400\ sec^{-100}$) pedestrians, respectively. We stopped the simulation when all the N agents had crossed the road or when the time threshold had been reached, but considered a maximum of 10 pedestrians for each crossing event in order to compare simulations with our observed data. We set the number of simulations to 1000 for each hypothesis and for each set of tested parameters (Light colour * Country * gender * hypotheses, with the range of mimetic coefficients values for the mimetism hypothesis), or a total of 70 000 simulations.

g. Statistical analyses

In order to know whether departure time survival distributions follow exponential law (see Supplementary information) and the distributions of departure latencies according to pedestrian crossing rank follow parabolic law, linear regression was used to compare theoretical data to observed data with adjusted $R^2$. The best fitting distribution were chosen according to F-statistics. The Kolmogorov-Smirnov test was used to compare the distribution of simulated data to observed data. This test was revealed to be the best means to compare two distributions. In a Kolmogorov-Smirnov test, the higher the p-value, the lower the D-statistics, and the better the fitting. Analyses were performed in R 3.3.2, with α set at 0.05.

**Results**

1. Analyses of departure latencies according to pedestrian departure rank

The survival curves for departure latencies of following pedestrians at the green light ($y=0.7736*exp^{-0.029x}$, $R^2=0.98$, $F=88800$, $p<0.00001$) and waiting time at the red light both fit with exponential distributions ($y=1.0606*exp^{-0.0005x}$, $R^2=0.97$, $F=23540$, $p<0.00001$). The probability that a pedestrian will follow another at the green light is 0.029, a value that is about 22 and 17.5 times higher than the probability to be the first to leave the kerb in France and Japan, respectively (see Table 3). The probability that a pedestrian will follow another at the red light is 0.0005, i.e. about 7 and 16 times higher than the probability to be the first to leave the kerb in France and Japan, respectively. Several initial conclusions can be drawn from this result. First, the substantially higher probability indicates a mimetic process. Second, this mimetic process seems to differ according to the light colour and the country: French citizens show less mimetism at the red light compared to the green light, but more cases of first pedestrians crossing at the red light in France compared to Japan. However, the same mimetic process is observed for Japanese pedestrians crossing at the red light and the green light, with much fewer pedestrians crossing first at the red light.

We then analysed whether the departure latencies according to the pedestrian departure rank followed a parabolic curve, which indicates that a mimetic process underlies the decision to cross. Results are given in Table 3. Whilst mimetic process is seen to govern the behaviour of Japanese pedestrians, this mimetism does not seem to be as consistent in the behaviour of French pedestrians.

2. Comparison between observed crossing and simulated crossing

The modelling of crossing behaviours reveals whether an individual bases their crossing decision purely on their own motivation (independence hypothesis) or if this decision solely relies on mimetic behaviours (mimetic process hypothesis).

*Crossings at the green light:* Only the number of pedestrians that were already crossing was used, and this was sufficient to explain the distribution of observed departure latencies (Table 3). Results are illustrated in Figure 1. No confirmation of the independence hypothesis was found for either country or gender, and results from simulations were consistently different from observational data ($p \leq 0.00001$, $D=1$). Values of mimetic coefficients for French men ranged from 0.0005 to 0.001 ($p \geq 0.126$, $D \leq 0.556$) and did not provide different distributions of simulated departure latencies compared to observed data with a best mimetic coefficient of 0.0006 ($p=0.352$, $D=0.444$, Fig. 1a). Concerning French women, distributions of simulated departure latencies with mimetic coefficients ranging from 0.0009 to 0.0015 were not statistically different from the observed data ($p \geq 0.126$,

D≤0.556) with a best mimetic coefficient of 0.00012 (p=0.730, D=0.333, Fig. 1b). Simulations for Japanese men showed similar distributions of departures latencies to observational data, with mimetic coefficients going from 0.0015 to 0.0019 (p ≥0.126, D≤0.556) with a best value of 0.0016 (p=0.989, D=0.222, Fig. 1c). Finally, simulations of crossings for Japanese woman are not different from observed data, with mimetic coefficients ranging from 0.0012 to 0.0015 (p ≥0.126, D≤0.556) and a best mimetic coefficient of 0.00135 (p=0.989, D=0.222, Fig. 1d).

*Crossings at the red light*: Although the number of pedestrians already crossing was sufficient to explain the departure latencies of French sites, the departure probability had to be modelled as dependent on $\frac{\text{the number of already crossing pedestrians}}{\text{the number of waiting pedestrians}}$ in order to fit the simulations to observation data in Japanese sites (Table 3). While all the waiting agents crossed in the green light model, some simulations for the red light model were different, with the number of crossing agents sometimes stopping before the threshold of ten was reached. This result was important to explain the data. The independence hypothesis was not confirmed, either because the distribution of simulated departure latencies was different from the observed data (p≤0.00001, D=1 for French pedestrians), or because the simulations only reached one follower in Japanese pedestrians, which is far from the result we obtained with observations (nine followers for Japanese men and seven followers for Japanese women). For French men, departure latencies of mimetic coefficients [0.0006; 0.0005] were not statistically different from observed departure latencies (p ≥0.126, D≤0.556; 0.0006 being the best mimetic coefficient: p=0.352, D=0.444, Fig. 2a). Concerning French women, distributions of simulated departure latencies with mimetic coefficients going from 0.0006 to 0.00085 are not statistically different from observed data (p ≥0.126, D≤0.556), with a best mimetic coefficient of 0.00085 (p=0.730, D=0.333, Fig. 2b). As previously noted, the number of pedestrians already crossing was not enough as a single process in our model to explain the departure latencies of Japanese pedestrians (p<0.028, D>0.722). However, the distributions of departure latencies in simulations that include the number of pedestrians already crossing and the number of waiting pedestrians (Table 3) are not different from the observed departure latencies (p ≥0.108, D≤0.667), with 0.005 as the best mimetic coefficient for Japanese men (p=0.860, D=0.264, Fig. 2c) and 0.0035 for Japanese woman (p=0.872, D=0.285, Fig. 2d).

Finally, for the rate of illegal crossings in simulations, the mimetism hypothesis –taking the number of crossing pedestrians into account for French sites and the number of crossing and waiting pedestrians into account for the Japanese sites – is the only hypothesis that provides similar results to observed rule breaking (Table 4). The simulated rate is the same as for the observed rate in all groups except French women, where we have 12% less rule breaking in observation data.

**Discussion**

This study showed that modelling is useful to test different hypotheses and find the exact processes underlying decision-making when pedestrians cross the road. It identified exactly which parameters are important to fit the simulations to observations, and revealed the differences between the Japanese and the French and between women and men in their decision to follow and cross at the green and red light. Finally, the model highlighted not only similar departure latencies but also a similar number of crossing pedestrians and a similar rate of illegal crossings between simulated and observed data.

In order to fit the simulated departure latencies to the observed departure latencies, we had to include different parameters in the model, certain of which varied according to the gender and the country. It is important to make the agent-based model reflect the site variables as closely as possible to be sure that simulation results are comparable to empirical observation data [36,53,54]. This avoids false positives or false negatives that can occur though incorrect model parameters. First, the number of waiting pedestrians was implemented. The distribution was identical for French and Japanese sites, confirming that our sites were comparable (as Table 1 shows), and that the number of waiting pedestrians was a necessary variable in the model to fit the simulated number of crossing pedestrians to the observed data. We had to model the threshold time rules of crossing that we observed according to the light colour but also according to the country. This threshold time necessity in the model confirms that decision making differs according to the light colour but also according to the country, with no crossing at the red light for Japanese pedestrians until the light for vehicles turns orange (see Part 1, Pelé et al. submitted). This effect of traffic light sequences has already been described in different studies [20,55]. Finally, we had to implement the intrinsic probability to cross and the mimetic coefficient to differentiate between the independence hypothesis and the mimetic process hypothesis, respectively.

The intrinsic probability, meaning our own motivation without being influenced by others, was revealed to differ according to the parameters we studied. First, the intrinsic probability is higher at the green light than at the red light, which is both understandable and reassuring. This shows that the probability of deciding to cross based solely on our personal information is lower at the red light compared to the green light. As we found in a previous study ([15]; and Part 1, Pelé et al. submitted), French pedestrians cross at the red light more often than Japanese pedestrians. The current study finds that the intrinsic probability to cross at the red light is higher at French sites and conversely, the intrinsic probability to cross at the green light is higher at Japanese sites. This difference is however not just related to the rate of illegal crossings, but is also explained by departure times. Japanese pedestrians start crossing sooner than their French counterparts at the green light, probably because they are more attentive to the light change, meaning that they trust their personal information more than they trust social information [12,15]. In stark contrast, the French cross more and much faster at the red light. Intrinsic probability is therefore representative of individual risk-taking levels [10,30,32]. We found a slight difference between French men and women crossing at the red light: here, women seem to have a lower probability of crossing based solely on their personal information at the red light, which confirms previous studies [11,26,56]. However, this intrinsic probability alone was not enough to explain the following behaviours of pedestrians, meaning that a mimetic or amplification process underlies most of the crossing decisions made by following pedestrians.

The implementing of a mimetic process in this model makes it possible to fit simulated data to empirical data and to explain them: the higher the number of individuals crossing, whatever the light colour, the higher the probability is that other individuals will cross. This copying behaviour, referred to as "sheep" or "herd" behaviour, is well known in human beings in domains such as financial markets, fashion, purchasing or crowd behaviour [5,57–61]. This process seems to be deeply rooted in human behaviour due to their gregariousness and sociality, and can be observed in many other social animal species [62–64]. Mimetism is lower for French pedestrians crossing at the red light, but they only take the number of crossing pedestrians into consideration when making their decision. Contrarily, Japanese pedestrians show greater levels of mimetism at the red light, but they take both the number of pedestrians already crossing and the number of waiting pedestrians into consideration when taking their decision to cross. These results confirm the more individualistic nature of French society and the more collectivist and conformist behaviour in Japan [11,65,66]. Indeed, the different processes we observed in this study show that the Japanese consider the behaviour of other pedestrians: the Japanese tendency to take the behaviour of others into account could be explained by their fear of being criticized (social credibility), which is greater than the fear of being fined (risk exposure) [10,24]. Taking into account the number of crossing and waiting pedestrians is a "following the majority" rule that illustrate how humans adapt their individual behaviour to that of others. Indeed, this rule has been well described in animal species [44,62,63] including strongly gregarious species such as sheep [43]. The lower mimetic coefficients suggest that men are more individualistic than women, but this is only confirmed in France, whilst Japanese men and women behave in the same way.

This study highlighted that only two human variables, namely intrinsic probability and particularly mimetism, were sufficient and adequate to explain the departure latencies of pedestrians but also the rate of rule breaking. Matching values for each gender and each country led to a significant fitting between simulated and observed data. This leads us to conclude that when crossing the road, at least at signalised crossings, human beings behave like sheep, and that a high rate of accidents with cars might be due to this herding behaviour and the misuse of social information. A sound signal produced when pedestrians cross at the red light could be a solution to stop the first pedestrian from crossing but also to warn other pedestrians that are ready to follow that this is not the right time. To conclude, these studies about the decision-making processes of pedestrians during road-crossing are useful tools to conceive new safety and public education solutions in transport research.

**Acknowledgements**

We thank Caroline Bellut, Elise Debergue, Charlotte Gauvin, Anne Jeanneret, Thibault Leclere, Lucie Nicolas, Florence Pontier and Diorne Zausa for their help collecting data. We are grateful to Kunio Watanabe and Hanya Goro (Primate Research Institute, Kyoto University) for their help in authorizing data collection. We thank Joanna Munro from Munro Language Services for the English language editing.

**Author contributions statements**

MP and CS scored the data. MP, JLD and CS analysed the data. MP and CS wrote the manuscript. MP, JLD and CS reviewed the manuscript.

**Tables**

Table 1: Information about the studied sites in France and in Japan. Road-crossing speed was estimated by scoring the crossing speed of 20 random pedestrians for each site.

| | France - Strasbourg |
|---|---|

| Sites | Train Station | Pont des Corbeaux | Place Broglie | |
|---|---|---|---|---|
| Coordinates | 48.584474, 7.736135 | 48.579509, 7.750745 | 48.584559, 7.748628 | |
| Lanes | 2*1 | 2*2 | 2*1 | |
| Mean pedestrian flow per hour | 667 | 612 | 850 | |
| Mean road-crossing speed | 0.96±0.05 | 1.11±0.29 | 1.01±0.16 | |
| Data collection dates | 02/07-07/07/2014 | 01/10-25/10/2014 | 15/02-09/03/2015 | |

| Japan - Nagoya | | | | |
|---|---|---|---|---|
| Sites | Train Station | Maruei | Excelco | Osu-Kannon |
| Coordinates | 35.170824, 136.884328 | 35.168638, 136.905740 | 35.166891, 136.907284 | 35.159316, 136.901697 |
| Lanes | 2*3 | 1*1 | 2*1 | 2*1 |
| Mean pedestrian flow per hour | 480 | 645 | 869 | 814 |
| Mean road-crossing speed | 1.10±0.22 | 1.15±0.21 | 0.98±0.21 | 1.07±0.18 |
| Data collection dates | 13/06-05/07/2011 | | 27/01-05/02/2015 | |

Table 2: Values of departure probability for the first pedestrian to leave the kerb $\psi_{01}$, intrinsic departure probability of following pedestrians $\lambda$, mimetic coefficient C and crossing rule according to the light colour, the country and the gender. Range of calculated C is given by resolving by parabolic equation (2c). The best mimetic coefficient C is the one obtained after implementing the range of calculated C in the model and confronting simulated and observed data. The crossing rule is considered to be that the probability to join depends on the number of already crossing pedestrians (equation 2a), or is calculated as the number of crossing pedestrians divided by the number of waiting ones (equation 3).

| Light colour | Country | Gender | $\psi_{01}$ | $\lambda$ | Range of calculated C | Best C | Crossing rule |
|---|---|---|---|---|---|---|---|
| **Green** | France | Man | 0.013 | 0.0013 | 0.0001-0.0026 | 0.0006 | Crossing |
| **Green** | France | Woman | 0.013 | 0.0013 | 0.0005-0.0031 | 0.0012 | Crossing |
| **Green** | Japan | Man | 0.016 | 0.0016 | 0.0006-0.0026 | 0.0016 | Crossing |
| **Green** | Japan | Woman | 0.017 | 0.0017 | 0.0004-0.0031 | 0.00135 | Crossing |
| **Red** | France | Man | 0.0008 | 0.00008 | 0.00047-0.0006 | 0.0005 | Crossing |
| **Red** | France | Woman | 0.0006 | 0.00006 | 0.00083-0.0002 | 0.0008 | Crossing |
| **Red** | Japan | Man | 0.0003 | 0.00003 | 0.0056-0.011 | 0.005 | Crossing/Waiting |
| **Red** | Japan | Woman | 0.0003 | 0.00003 | 0.0021-0.0032 | 0.005 | Crossing/Waiting |

Table 3: Statistical values of parabolic curve estimation test with the departure latencies according to the pedestrian departure rank, for each light colour, gender and country. Significant results are indicated in bold.

| | | Green light | | | Red light | | |
|---|---|---|---|---|---|---|---|
| | | df | F | p | df | F | p |
| **France** | Man | 7 | 2.79 | 0.139 | **7** | **15.37** | **0.006** |
| **France** | Woman | 7 | 4.01 | 0.08 | 7 | 1.448 | 0.268 |
| **Japan** | Man | **7** | **42.02** | **0.0003** | **7** | **5.93** | **0.026** |
| **Japan** | Woman | **7** | **40.1** | **0.0004** | **4** | **7.81** | **0.039** |

Table 4: Rate of illegal crossings for observations and simulations. C means model with only crossing pedestrian as rule, C/W means number of crossing pedestrians divided by number of waiting pedestrians rule.

| | France | | Japan | |
|---|---|---|---|---|
| | men | women | men | women |
| **Observed** | 0.46 | 0.38 | 0.02 | 0.02 |
| **Independence hyp.** | 0.24 | 0.19 | 0.0006 | 0.001 |
| **Mimetic hyp. (C)** | 0.50 | 0.50 | 0.12 | 0.04 |
| **Mimetic hyp. (C/W)** | NA | NA | 0.02 | 0.02 |

**Figure legends:**

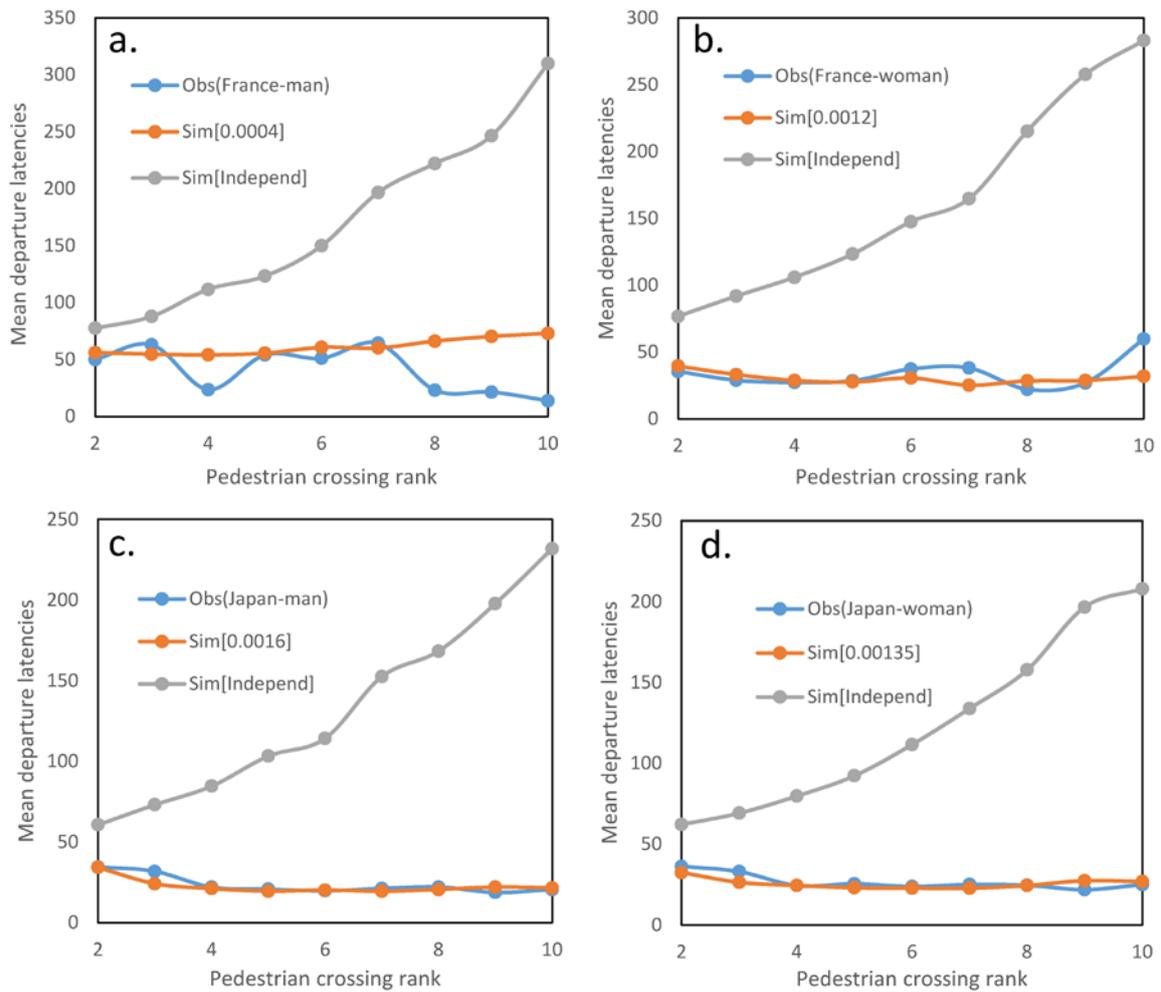

Figure 1: Distribution of mean departure latencies (sec$^{-100}$) at the green light according to pedestrian crossing rank for observed data (blue), simulated data of the independence hypothesis (grey) and simulated data for the best mimetic coefficient (orange). (a.) For French men, (b.) for French women, (c.) for Japanese men and (d.) for Japanese women.

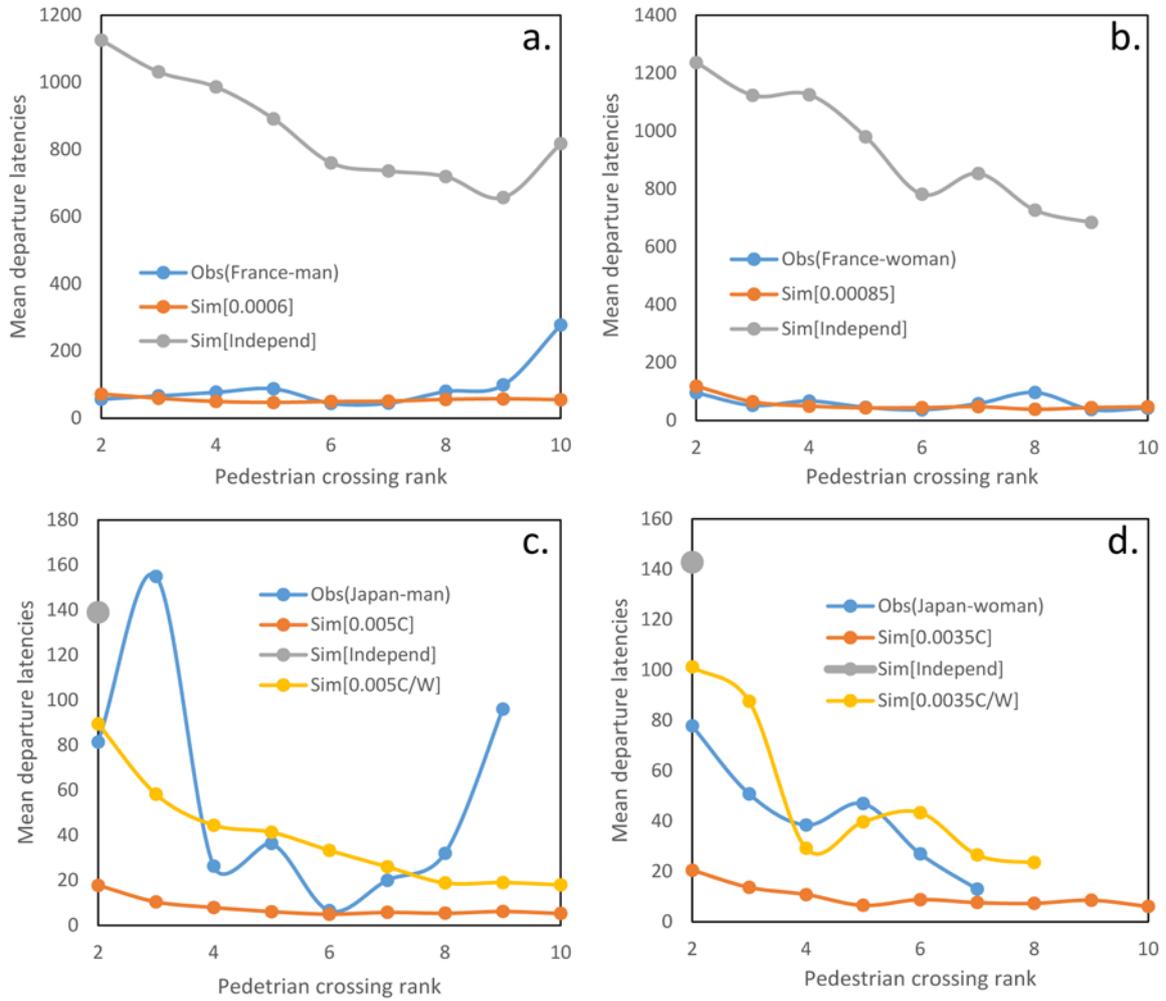

Figure 2: Distribution of mean departure latencies (sec$^{-100}$) at the red light according to pedestrian crossing rank for observed data (blue), simulated data of the independence hypothesis (grey) and simulated data for the best mimetic coefficient (orange, and solely involves pedestrians that are already crossing for Japanese sites). Finally, yellow curves indicate the ratio of crossing/waiting pedestrians for the Japanese sites. (a.) For French men, (b.) for French women, (c.) for Japanese men and (d.) for Japanese women.

**Supplementary information for**

**Decision-making processes underlying pedestrian behaviours at signalised crossings:**
**Part 2. Do pedestrians show cultural herding behaviour ?**

Marie Pelé[1], Jean-Louis Deneubourg[2], Cédric Sueur[3]

[1] Ethobiosciences, Research and Consultancy Agency in Animal Well-Being and Behaviour, Strasbourg, France

[2] Unit of Social Ecology, Université libre de Bruxelles, Brussels, Belgium

[3] Université de Strasbourg, CNRS, IPHC UMR 7178, F-67000 Strasbourg, France

Corresponding author: marie.pele@iphc.cnrs.fr, 0033(0)88107453, IPHC UMR 7178, 23 rue Becquerel F-67000 Strasbourg, France

Here are given the calculation of the mimetic coefficient for each gender and country, at the green and at the red light

1. **Green light**

    a. **France man**

$\lambda = 0.013/n$, n=10, then $\lambda = 0.0013$ (FigS1)

$$\frac{1}{\Delta T_{j-1,j}} = -0.0153 + 0.0003\,j - 0.0005\,j^2 \qquad \text{(FigS2)}$$

then

$$(\lambda - C)(n+1) + j(2C + Cn - \lambda) - Cj^2 = -0.0153 + 0.0003\,j - 0.0005\,j^2$$

So $(\lambda - C)(n+1) = -0.0153$, $(2C + Cn - \lambda) = 0.0003$ and $C = 0.0005$

$$C = -\left(\frac{-0.0153}{(n+1)} - \lambda\right) = 0.0026$$

or $C = \dfrac{0.0003 + \lambda}{2 + n} = 0.0001$

or $C = 0.0005$

then $C$ approximately equals 0.0001-0.0026

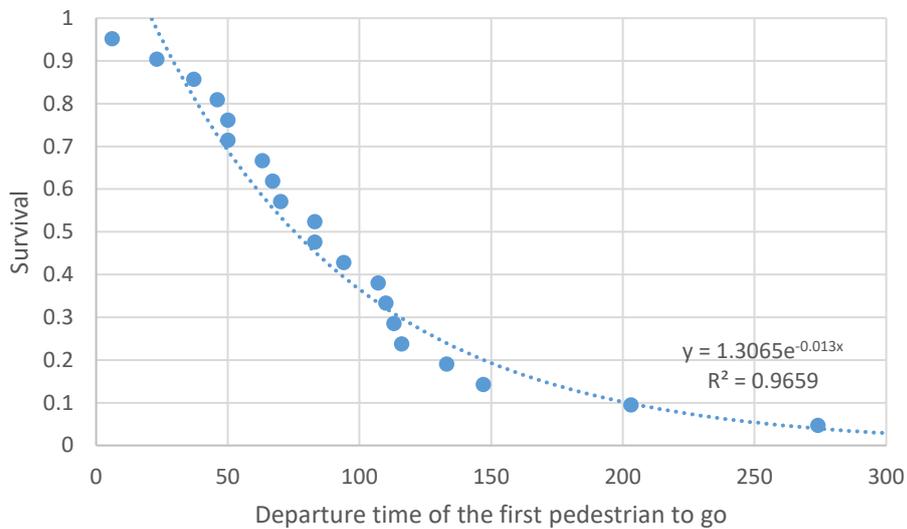

Figure S1: Survival analysis of departure time of the first pedestrian to go for French men at the green light. The survival follows an exponential law (df=18, F=228, p<0.00001).

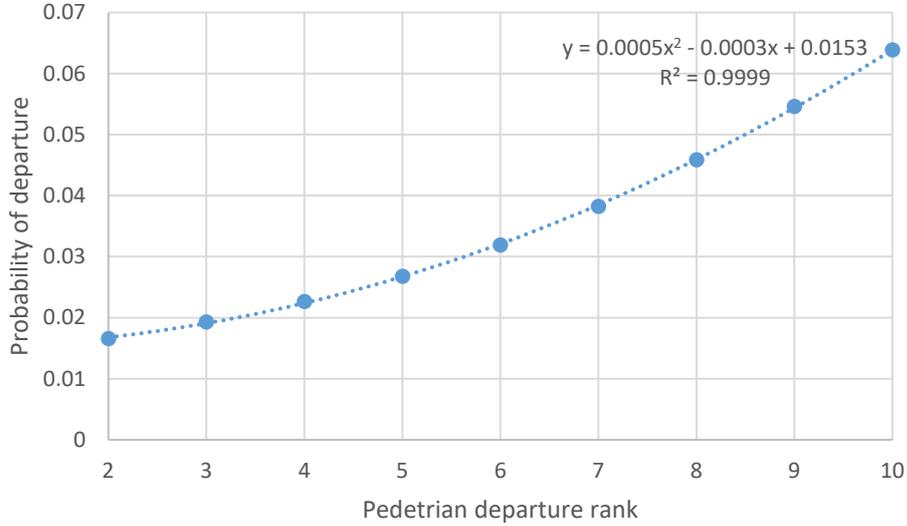

Figure S2: Probability of departure according to pedestrian departure rank for French men at the green light.

**b. France woman**

$\lambda$ =0.013/n, n=10, then $\lambda$ = 0.0013 (FigS3)

$$\frac{1}{\Delta T_{j-1,j}} = -0.0198 + 0.0057\,j - 0.0005\,j^2 \qquad \text{(FigS4)}$$

then

$$(\lambda - C)(n+1) + j(2C + Cn - \lambda) - Cj^2 = -0.0198 + 0.0057\,j - 0.0005\,j^2$$

So $(\lambda - C)(n+1) = -0.0198$, $(2C + Cn - \lambda) = 0.0057$ and $C = 0.0005$

$$C = -\left(\frac{-0.0198}{(n+1)} - \lambda\right) = 0.0031$$

or $C = \dfrac{0.0057 + \lambda}{2 + n} = 0.00058$

or $C = 0.0005$

then $C$ approximately equals 0.0005-0.0031

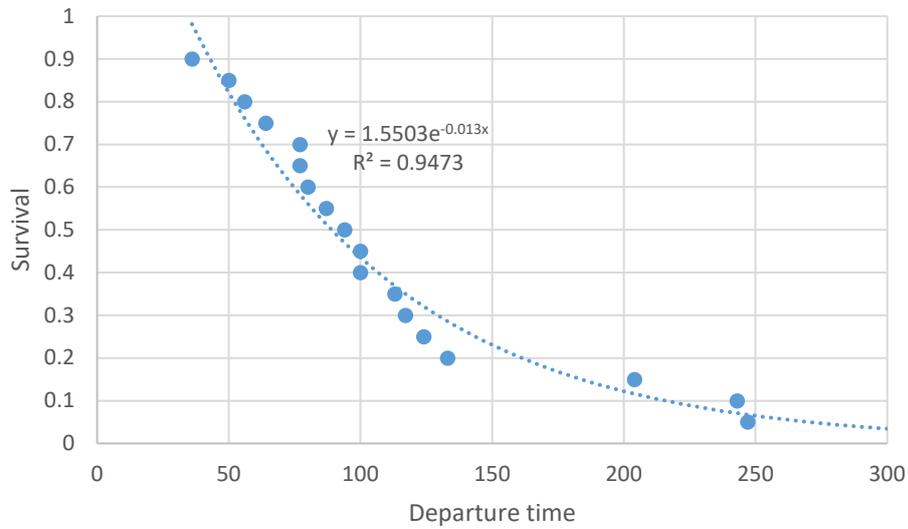

Figure S3: Survival analysis of departure time of the first pedestrian to go for French women at the green light. The survival follows an exponential law (df=18, F=93, p<0.00001).

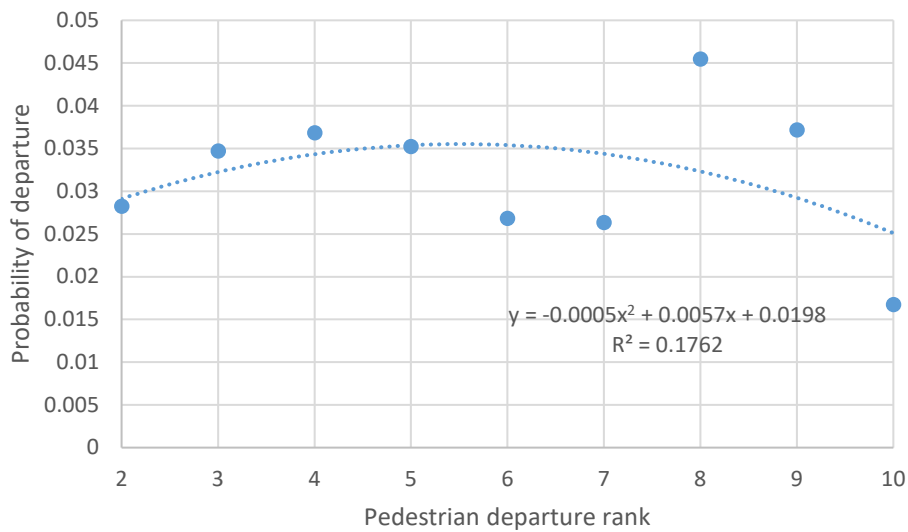

Figure S4: Probability of departure according to pedestrian departure rank for French women at the green light.

### c. **Japan man**

$\lambda = 0.016/n$, n=10, , then $\lambda = 0.0016$ (FigS5)

$$\frac{1}{\Delta T_{j-1,j}} = -0.0113 + 0.01j - 0.0006j^2 \qquad \text{(FigS6)}$$

then

$$(\lambda - C)(n+1) + j(2C + Cn - \lambda) - Cj^2 = -0.0113 + 0.01j - 0.0006j^2$$

So $(\lambda - C)(n+1) = -0.0113$, $(2C + Cn - \lambda) = 0.01$ and $C = 0.0006$

$$C = -\left(\frac{-0.0113}{(n+1)} - \lambda\right) = 0.0026$$

or $C = \dfrac{0.01 + \lambda}{2 + n} = 0.00096$

or $C = 0.0006$

then $C$ approximately equals 0.0006-0.0026

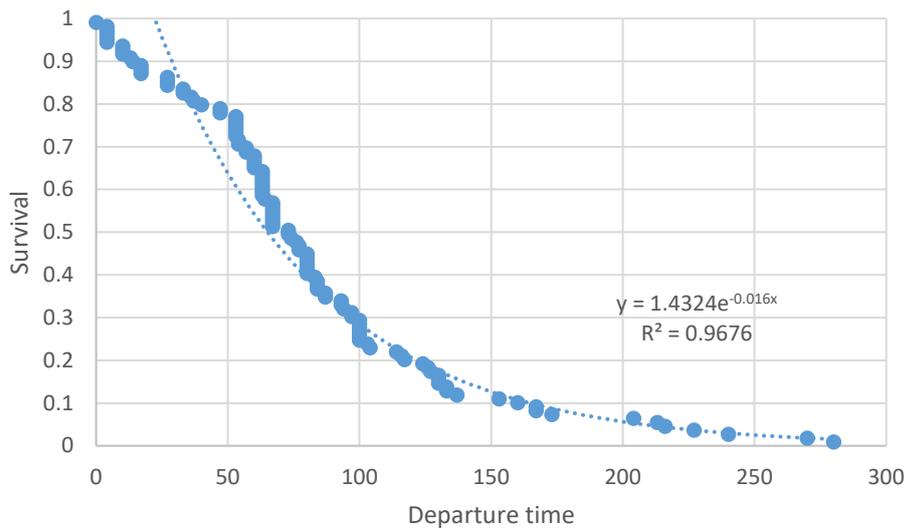

Figure S5: Survival analysis of departure time of the first pedestrian to go for Japanese men at the green light. The survival follows an exponential law (df=106, F=772, p<0.00001).

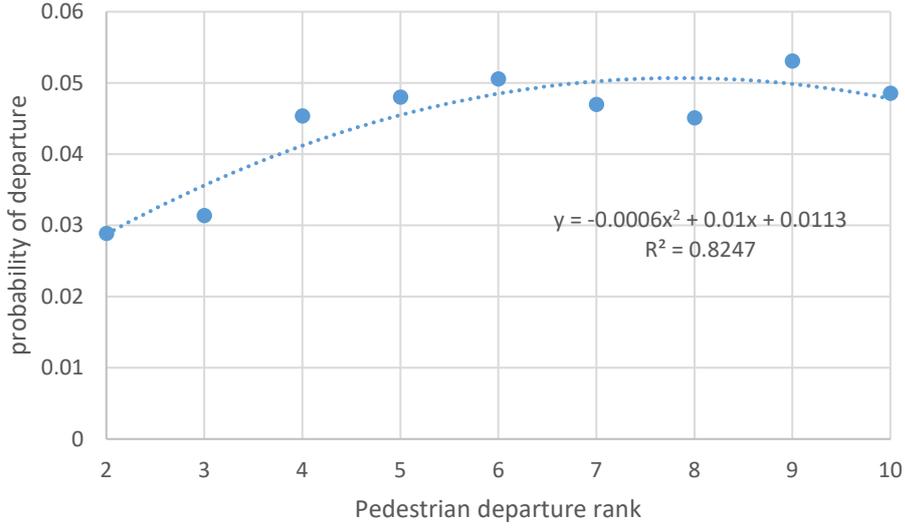

Figure S6: Probability of departure according to pedestrian departure rank for Japanese men at the green light.

### d. __Japan woman__

λ =0.017/n, n=10, , then λ = 0.0017 (FigS7)

$$\frac{1}{\Delta T_{j-1,j}} = -0.0157 + 0.0069\,j - 0.0004\,j^2 \qquad \text{(FigS8)}$$

then

$$(\lambda - C)(n+1) + j(2C + Cn - \lambda) - Cj^2 = -0.0157 + 0.0069\,j - 0.0004\,j^2$$

So $(\lambda - C)(n+1) = -0.0157$ , $(2C + Cn - \lambda) = 0.0069$ and $C = 0.0004$

$$C = -\left(\frac{-0.0157}{(n+1)} - \lambda\right) = 0.0031$$

or $C = \dfrac{0.0069 + \lambda}{2+n} = 0.0007$

or $C = 0.0004$

then $C$ approximately equals 0.0004-0.0031

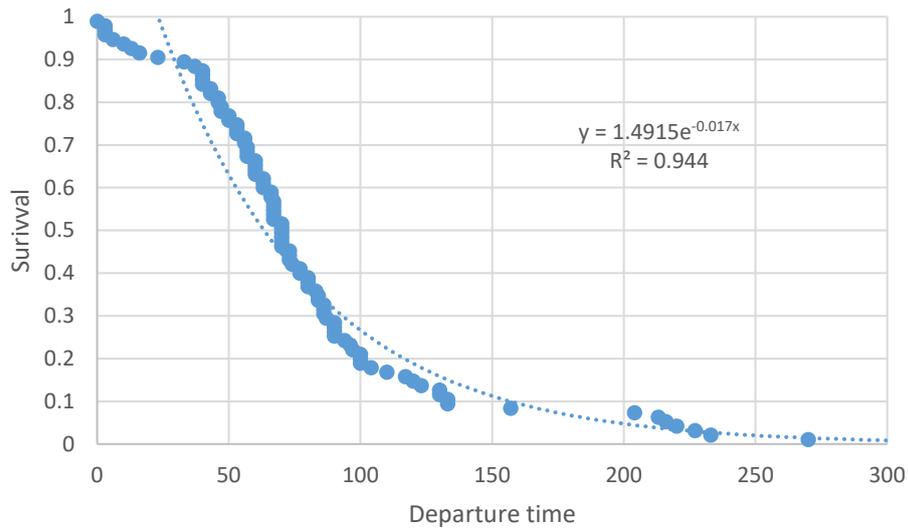

Figure S7: Survival analysis of departure time of the first pedestrian to go for Japanese women at the green light. The survival follows an exponential law (df=93, F=452, p<0.00001).

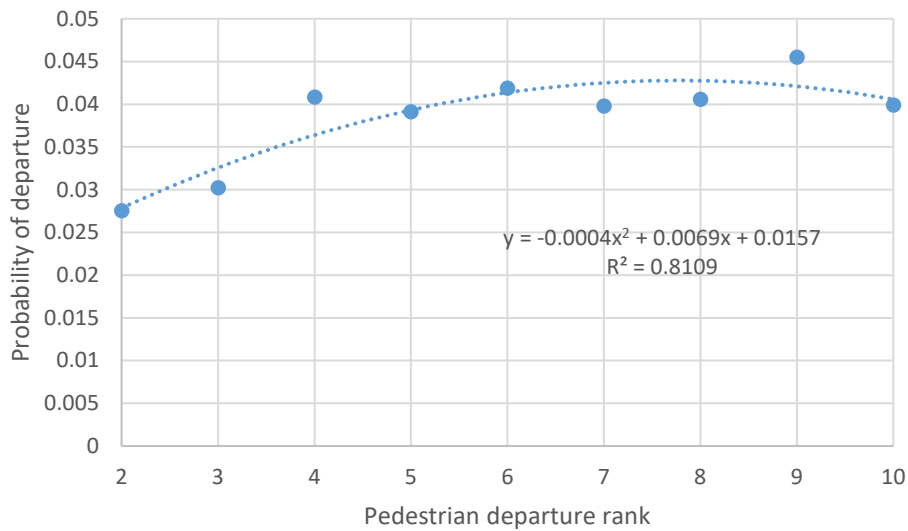

Figure S8: Probability of departure according to pedestrian departure rank for Japanese women at the green light.

2. **Red light**

    a. **France man**

$\lambda = 0.0008/n$, $n=10$, , then $\lambda = 0.00008$ (FigS9)

$$\frac{1}{\Delta T_{j-1,j}} = -0.0058 + 0.0049\,j - 0.0005\,j^2 \qquad \text{(FigS10)}$$

then

$$(\lambda - C)(n+1) + j(2C + Cn - \lambda) - Cj^2 = -0.0058 + 0.0049\,j - 0.0005\,j^2$$

So $(\lambda - C)(n+1) = -0.0058$, $(2C + Cn - \lambda) = 0.0049$ and $C = 0.0005$

$$C = -\left(\frac{-0.0058}{(n+1)} - \lambda\right) = 0.0006$$

or $C = \dfrac{0.0049 + \lambda}{2 + n} = 0.000475$

or $C = 0.0005$

then $C$ approximately equals 0.000475-0.0006

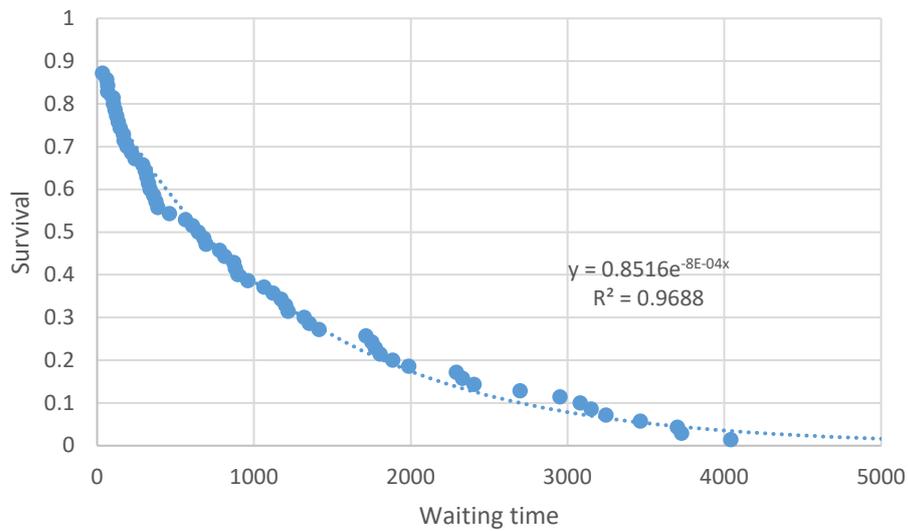

Figure S9: Survival analysis of waiting time of the first pedestrian to go for French men at the red light. The survival follows an exponential law (df=68, F=3884, p<0.00001).

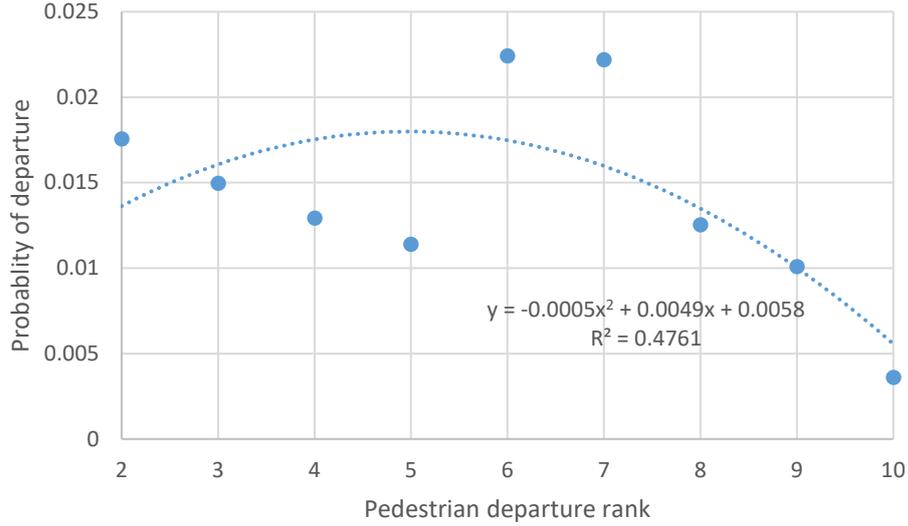

Figure S10: Probability of departure according to pedestrian departure rank for French men at the red light.

### b. __France woman__

$\lambda =0.0006/n$, n=10, , then $\lambda = 0.00006$ (FigS11)

$$\frac{1}{\Delta T_{j-1,j}} = -0.0085 + 0.0028\,j - 0.0002\,j^2$$

(FigS12)

then

$$(\lambda - C)(n+1) + j(2C + Cn - \lambda) - Cj^2 = -0.0085 + 0.0028\,j - 0.0002\,j^2$$

So $(\lambda - C)(n+1) = -0.0085$, $(2C + Cn - \lambda) = 0.0028$ and $C = 0.0002$

$$C = -\left(\frac{-0.0085}{(n+1)} - \lambda\right) = 0.00083$$

or $C = \dfrac{0.0028 + \lambda}{2 + n} = 0.00024$

or $C = 0.0002$

then $C$ approximately equals 0.00083-0.0002

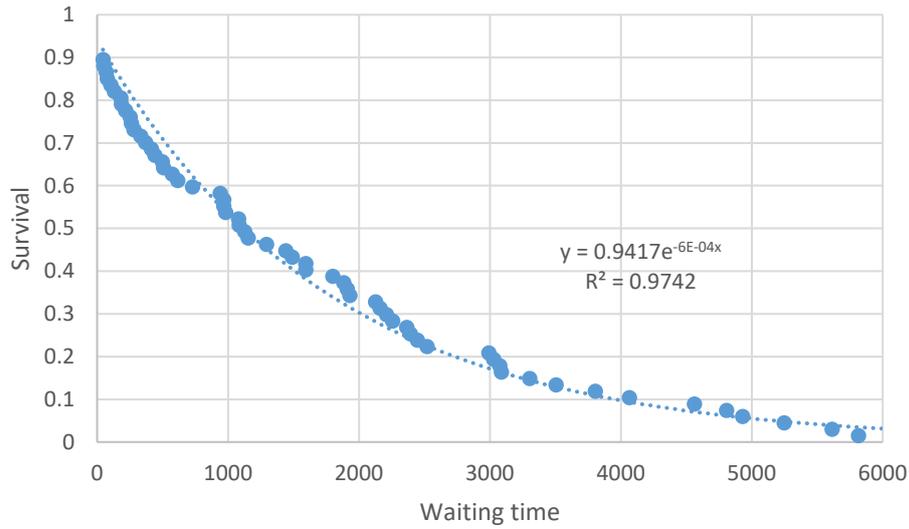

Figure S11: Survival analysis of waiting time of the first pedestrian to go for French women at the red light. The survival follows an exponential law (df=65, F=5606, p<0.00001).

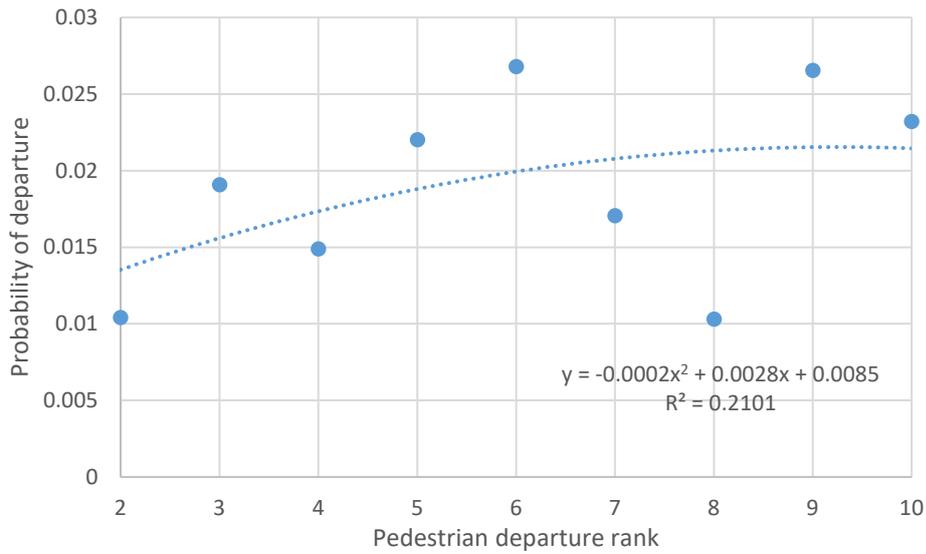

Figure S12: Probability of departure according to pedestrian departure rank for French women at the red light.

### c. **Japan man**

$\lambda$ =0.0003/n, n=10, , then $\lambda$ = 0.00003 (FigS13)

$$\frac{1}{\Delta T_{j-1,j}} = -0.1216 + 0.067\,j - 0.0058\,j^2$$

(FigS14)

then

$$(\lambda - C)(n+1) + j(2C + Cn - \lambda) - Cj^2 = -0.1216 + 0.067\,j - 0.0058\,j^2$$

So $(\lambda - C)(n+1) = -0.1216$, $(2C + Cn - \lambda) = 0.067$ and $C = 0.0058$

$$C = -\left(\frac{-0.1216}{(n+1)} - \lambda\right) = 0.011$$

or $C = \dfrac{0.067 + \lambda}{2 + n} = 0.0056$

or $C = 0.0005$

then $C$ approximately equals 0.0056-0.011

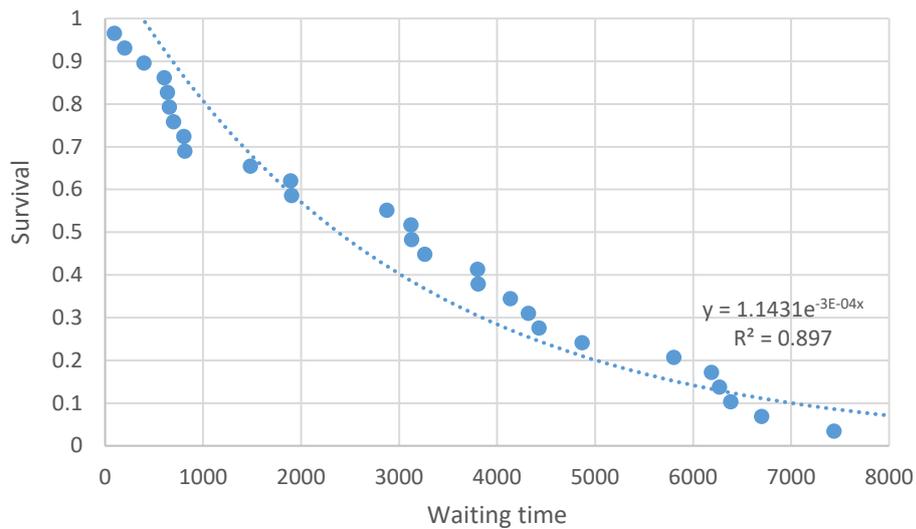

Figure S13: Survival analysis of waiting time of the first pedestrian to go for Japanese men at the red light. The survival follows an exponential law (df=27, F=627, p<0.00001).

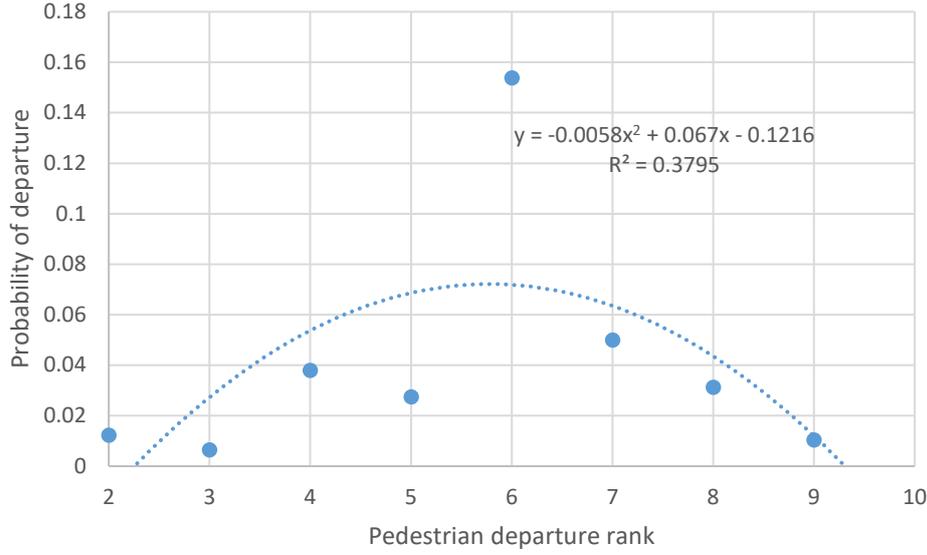

Figure S14: Probability of departure according to pedestrian departure rank for Japanese men at the red light.

### d. **Japan woman**

$\lambda = 0.0003/n$, $n=10$, , then $\lambda = 0.00003$ (FigS15)

$$\frac{1}{\Delta T_{j-1,j}} = -0.035 + 0.0255\,j - 0.0021\,j^2 \qquad \text{(FigS16)}$$

then

$$(\lambda - C)(n+1) + j(2C + Cn - \lambda) - Cj^2 = -0.035 + 0.0255\,j - 0.0021\,j^2$$

So $(\lambda - C)(n+1) = -0.035$, $(2C + Cn - \lambda) = 0.0255$ and $C = 0.0021$

$$C = -\left(\frac{-0.035}{(n+1)} - \lambda\right) = 0.0032$$

or $C = \dfrac{0.0255 + \lambda}{2 + n} = 0.0021$

or $C = 0.0021$

then $C$ approximately equals 0.0021-0.0032

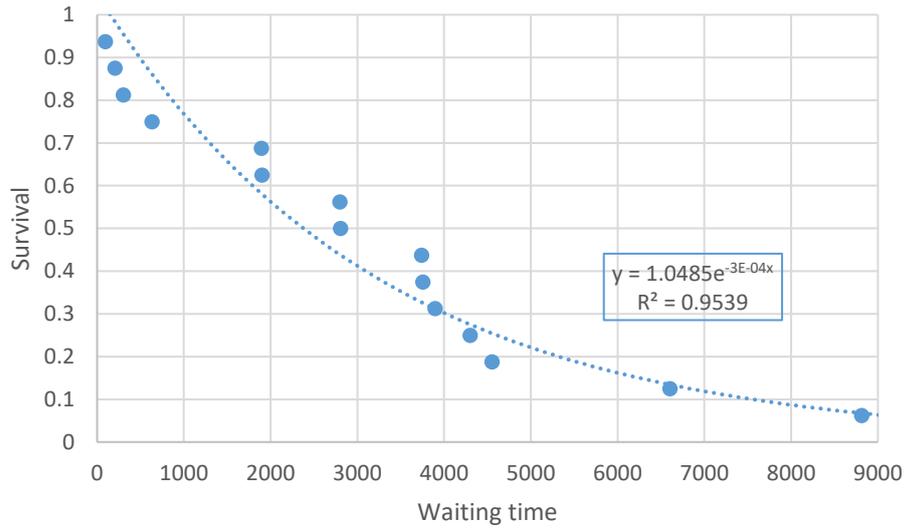

Figure S15: Survival analysis of waiting time of the first pedestrian to go for Japanese women at the red light. The survival follows an exponential law (df=14, F=220, p<0.00001).

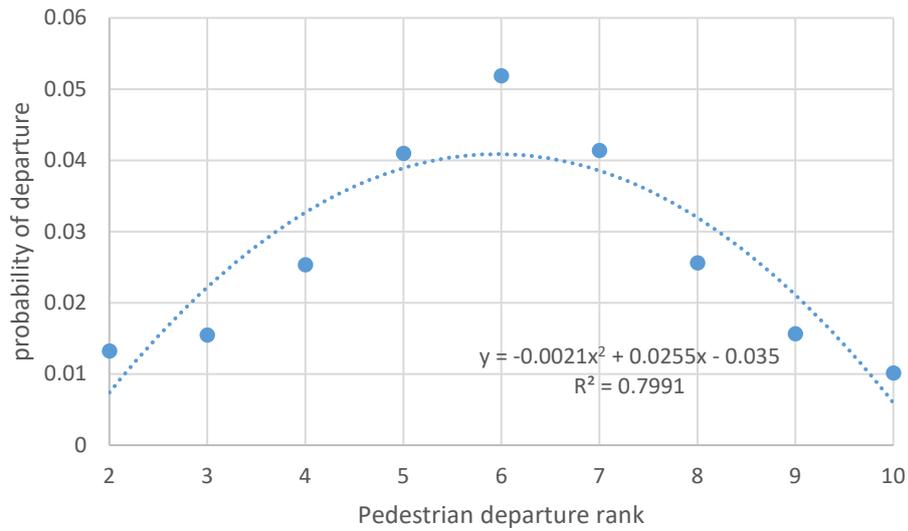

Figure S14: Probability of departure according to pedestrian departure rank for Japanese women at the red light.